\documentclass{Interspeech2024}

\interspeechcameraready

\usepackage[nolist]{acronym}
\usepackage{subcaption}
\usepackage{graphicx}
\usepackage{tabularx}
\usepackage{booktabs}
\usepackage{multirow}
\usepackage{pifont}
\usepackage{csquotes}
\newcommand{\cmark}{\ding{51}}%
\newcommand{\xmark}{\ding{55}}%
\newcolumntype{R}{>{\raggedleft\arraybackslash}X}
\usepackage{url}[hyphens]

\usepackage{hyphenat}
\begin{acronym}
\acro{DFT}{discrete Fourier transform}
\acro{MVDR}{minimum variance distortionless response}
\acro{PDF}{probability density function}
\acro{MMSE}{minimum mean square error}
\acro{ML}{maximum likelihood}
\acro{RIR}{room impulse response}
\acro{SNR}{signal-to-noise ratio}
\acro{SI-SDR}{scale-invariant signal-to-distortion ratio}
\acro{SDR}{signal-to-distortion ratio}
\acro{SIR}{signal-to-interference ratio}
\acro{SAR}{signal-to-artifact ratio}
\acro{ESTOI}{extended short-time objective intelligibility}
\acro{MAP}{maximum a posteriori}
\acro{ASR}{automatic speech recognition}
\acro{POLQA}{perceptual objective listening quality analysis}
\acro{MOS}{mean opinion score}
\acro{PESQ}{perceptual evaluation of speech quality}
\acro{EM}{expectation maximization}
\acro{STFT}{short-time Fourier transform}
\acro{DNN}{deep neural network}
\acro{SE}{speech enhancement}
\acro{VAE}{variational autoencoder}
\acro{GAN}{generative adversarial network}
\acro{SDE}{stochastic differential equation}
\acro{ELBO}{evidence lower bound}
\acro{RIR}{room impulse response}
\acro{DRR}{direct-to-reverberant ratio}
\acro{SSL}{self-supervised learning}
\acro{NCSN++}{Noise Conditional Score Network}
\acro{DNN}{deep neural network}
\acro{LKFS}{loudness K-weighted relative to full scale}
\acro{MAC}{multiply–accumulate operation}
\acro{GPU}{graphics processing unit}
\acro{MOS}{mean opinion scores}
\acro{SI-SDR}{scale-invariant signal-to-distortion ratio}
\acro{WER}{word error rate}
\end{acronym}

\title{EARS: An Anechoic Fullband Speech Dataset Benchmarked for \break Speech Enhancement and Dereverberation}

\name[affiliation={1}]{Julius}{Richter}
\name[affiliation={2}]{Yi-Chiao}{Wu}
\name[affiliation={2}]{Steven}{Krenn}
\name[affiliation={1}]{Simon}{Welker}
\name[affiliation={1}]{Bunlong}{Lay}
\name[affiliation={3}]{Shinji}{Watanabe}
\name[affiliation={2}]{Alexander}{Richard}
\name[affiliation={1}]{Timo}{Gerkmann}

\address{
  $^1$Signal Processing (SP), Universität Hamburg, Germany\\
  $^2$Codec Avatars Lab, Meta, Pittsburgh, PA, USA \\
  $^3$Language Technology Institute, Carnegie Mellon University, Pittsburgh, PA, USA}
\email{julius.richter@uni-hamburg.de, timo.gerkmann@uni-hamburg.de}


\begin{document}

\maketitle

\begin{abstract}

We release the EARS (\textbf{E}x\-pres\-sive \textbf{A}nechoic \textbf{R}ecordings of \textbf{S}peech)~dataset, a high-quality speech dataset comprising 107 speakers from diverse backgrounds, totaling in 100 hours of clean, anechoic speech data. The dataset covers a large range of different speaking styles, including emotional speech, different reading styles, non-verbal sounds, and conversational freeform speech. We benchmark various methods for speech enhancement and dereverberation on the dataset and evaluate their performance through a set of instrumental metrics. In addition, we conduct a listening test with 20 participants for the speech enhancement task, where a generative method is preferred. We introduce a blind test set that allows for automatic online evaluation of uploaded data. Dataset download links and automatic evaluation server can be found online\footnotemark[1].
\end{abstract}

\footnotetext[1]{\texttt{https://sp-uhh.github.io/ears\char95 dataset/}}

\section{Introduction}

Learning-based speech processing has seen huge leaps forward in recent years, with the impact of deep learning spanning essentially all areas from speech representation learning~\cite{mohamed2022self} over text-to-speech~\cite{tan2021survey} to speech enhancement~\cite{wang2018supervised}. Publicly available datasets such as LibriSpeech~\cite{panayotov2015librispeech} or VCTK~\cite{yamagishi2019vctk} have undoubtedly been a key driver of open and reproducible research in our field and have enabled steady progress. However, these datasets typically come with multiple shortcomings and are either too small, of low recording quality or do not span a large enough variety of different speakers and speaking styles.

To overcome these shortcomings, we release the \textbf{E}x\-pres\-sive \textbf{A}nechoic \textbf{R}e\-cord\-ings of \textbf{S}peech (EARS)~dataset. EARS contains 100$\,$h of anechoic speech recordings at 48$\,$kHz from over 100 English speakers with high demographic diversity. The dataset spans the full range of human speech, including reading tasks in seven different reading styles, emotional reading and freeform speech in 22 different emotions, conversational speech, and non-verbal sounds like laughter or coughing.

In addition, we set up a speech enhancement and speech dereverberation benchmark on EARS, comparing several predictive~\cite{luo2019conv, rouard2022hybrid} and generative~\cite{lu2022conditional, richter2023speech} speech enhancement methods. The benchmarks are intended to provide valuable insights into models' strengths, limitations, and comparability, thus promoting the development of robust and efficient speech enhancement systems. 

\begin{figure}[t]
    \centering
    \begin{subfigure}{0.21\linewidth}
        \includegraphics[width=\linewidth]{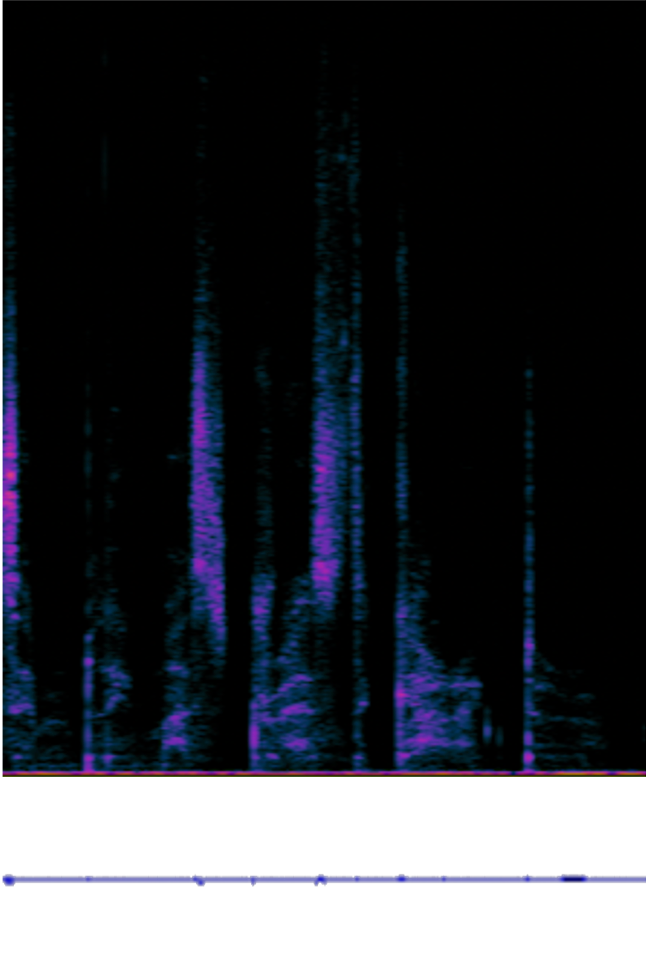}
        \caption{whisper}
        \label{fig:whisper}
    \end{subfigure}
    \begin{subfigure}{0.21\linewidth}
        \includegraphics[width=\linewidth]{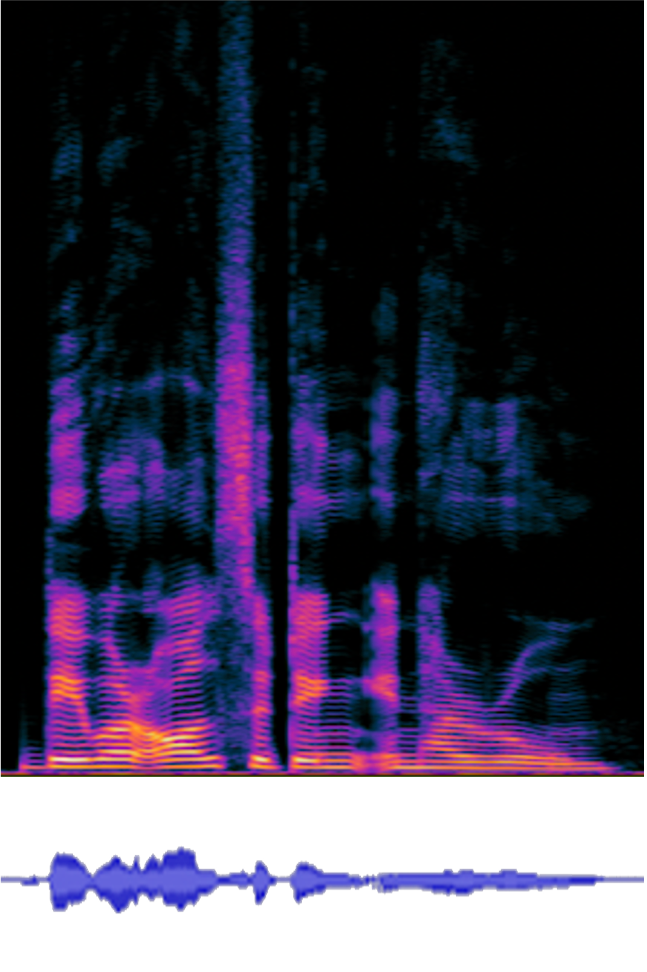}
        \caption{regular}
        \label{fig:regular}
    \end{subfigure}
    \begin{subfigure}{0.21\linewidth}
        \includegraphics[width=\linewidth]{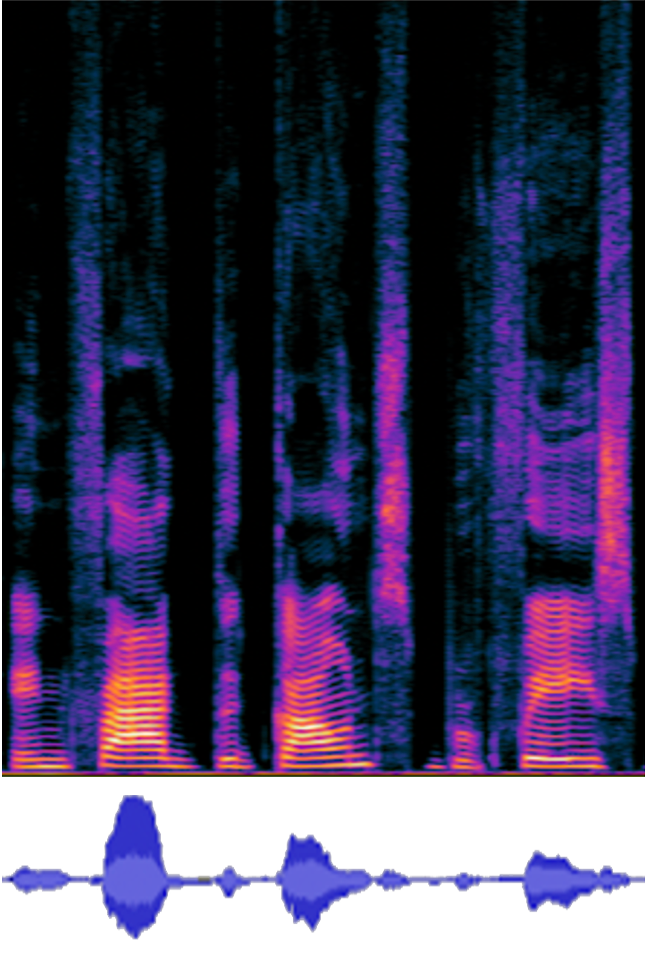}
        \caption{loud}
        \label{fig:loud}
    \end{subfigure}
    \begin{subfigure}{0.21\linewidth}
        \includegraphics[width=\linewidth]{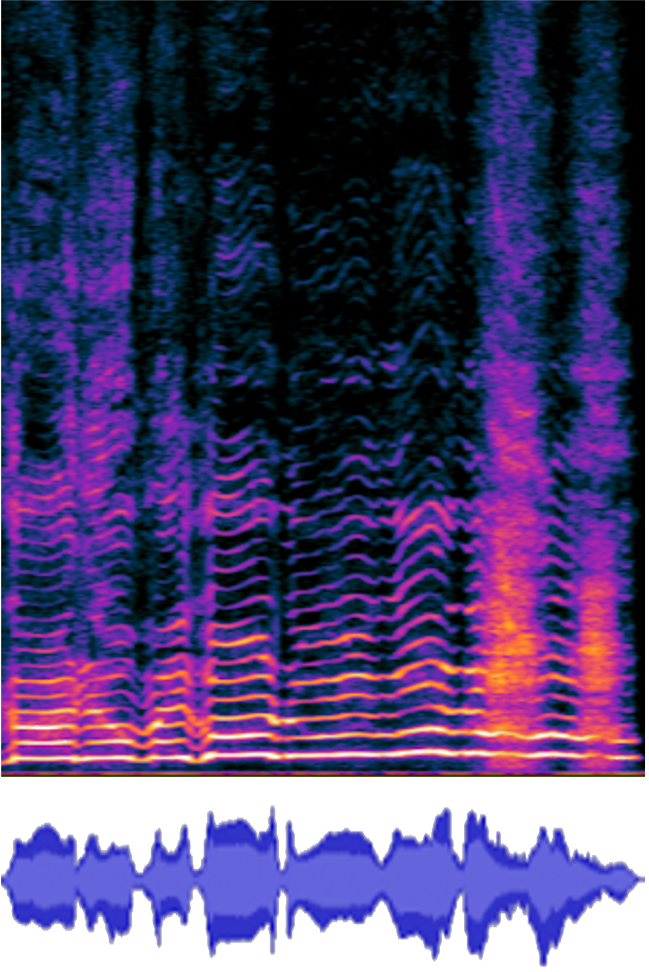}
        \caption{yelling}
        \label{fig:yelling}
    \end{subfigure}
    \begin{subfigure}{0.1198\linewidth}
        \includegraphics[width=\linewidth]{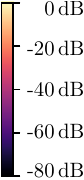}
            \vspace{1.14mm} 
        \caption*{}
    \end{subfigure}
    \caption{\textbf{High Dynamic Range.} The EARS~dataset spans the complete dynamic range of human speech, from whispering to yelling and screaming.}
    \label{fig:high_dynamic_range}
    \vspace{-1em}
\end{figure}

\begin{table*}[t]
    \footnotesize
    \centering
    \begin{tabularx}{1.0\linewidth}{@{}lRRRRRR>{\raggedleft\arraybackslash}p{1.8cm}R@{}}
        \toprule
    & hours & speakers & sample rate & anechoic & reading styles & freeform speech & emotional speech & speaker metadata \\ \midrule
    DNS (LibriVox)~\cite{dubey2023icassp}  & 556 & 1948 & 48$\,$kHz$^\dagger$ & \xmark & n/a & \xmark & \xmark &  \xmark \\
    MSP-Conversation (v1.0)~\cite{Martinez-Lucas_2020}  & 14 & 197 & 16$\,$kHz & \xmark & n/a & \cmark & \cmark & \xmark  \\
    MSP-Podcast (v1.10)~\cite{Lotfian_2019_3}  & 166 & 1458 & 16$\,$kHz & \xmark & n/a & \cmark & \cmark & \xmark  \\
    LibriSpeech~\cite{panayotov2015librispeech}  & 982 & 2484 & 16$\,$kHz & \xmark & n/a & \xmark & \xmark &  \xmark \\
    LJSpeech~\cite{ljspeech17} & 24 & 1 & 22.05$\,$kHz & \xmark & neutral & \xmark & \xmark & \xmark\\
    TIMIT~\cite{garofolo1993timit} & 5 & 632 & 16$\,$kHz & \xmark & neutral & \xmark & \xmark & \cmark \\
    VCTK~\cite{yamagishi2019vctk} & 44 & 110 & 48$\,$kHz & \xmark & neutral & \xmark & \xmark & \xmark \\
    WSJ0~\cite{garofolo1996wsj0} & 29 & 119 & 16$\,$kHz & \xmark & neutral & \xmark & \xmark & \xmark \\ \midrule
        EARS~(ours) & 100 & 107 & 48$\,$kHz & \cmark & 7 styles & \cmark & 22 emotions & \cmark \\
        \bottomrule
    \end{tabularx}
    \caption{\textbf{Speech datasets.} In contrast to existing datasets, the EARS dataset is of higher recording quality, large, and more diverse. Reading tasks feature seven styles (regular, loud, whisper, fast, slow, high pitch, and low pitch). Additionally, the dataset features unconstrained freeform speech and speech in 22 different emotional styles. We provide transcriptions of the reading portion and meta-data of the speakers (gender, age, race, first language). $^\dagger$contains files with limited bandwidth}
    \label{tab:dataset_comparisons}
    \vspace{-1em}
\end{table*}

\begin{table}[t]
\footnotesize
\begin{tabularx}{1.0\linewidth}{@{}lr>{\raggedright\arraybackslash}X@{}}
\toprule
 & \#$\,$files & rooms \\ \midrule
ACE-Challenge~\cite{eaton2016estimation} & 84 & building lobby, lecture room, meeting room, office \\
AIR~\cite{jeub2009binaural} & 344 & auditorium, corridor, lecture room, meeting room, stairway \\
ARNI~\cite{prawda2022robust} & 1000 & variable acoustics laboratory  \\
BRUDEX~\cite{fejgin2023brudex} & 144 & variable acoustics laboratory \\
dEchorate~\cite{carlo2021dechorate} & 648 & variable acoustics laboratory \\
DetmoldSRIR~\cite{amengual2020open} & 49 & concert hall, music chamber, theater \\ 
Palimpsest~\cite{Palimpsest} & 44 & air raid shelter, dockyard, submarine \\ \bottomrule
\end{tabularx}
\caption{\textbf{RIR datasets}. To construct EARS-Reverb, we use 2313 RIR files with different room characteristics.}
\label{tab:rir_datasets}
\vspace{-1em}
\end{table}

\section{EARS~dataset}

A good speech dataset is characterized by its scale, diversity, and high recording quality. However, most existing datasets fall short in one or more of these characteristics; see Table~\ref{tab:dataset_comparisons}. Most notably, a dataset that is of high recording quality (clean 48$\,$kHz audio), has a sufficient scale and covers the full range of human speech as opposed to only reading or neutral speech does not exist to the best of our knowledge. Yet, such a dataset is strongly required to advance research ranging from speech synthesis over voice and style conversion to speech enhancement.

We overcome these limitations with the EARS dataset, which provides high speaker and speech diversity paired with the highest recording quality.

\textbf{High Recording Quality.}
All speech is recorded in an anechoic chamber as 32-bit audio at 48$\,$kHz. We simultaneously record with a low-noise GRAS 40HH and a GRAS 48BL microphone, which are both mounted about 1$\,$m in front of the participant. The first microphone has low self-noise and high sensitivity to capture subtle and nuanced speech signals, while the second has lower sensitivity to capture high-energy speech like yelling without clipping, allowing us to capture the full dynamic range of human speech, see Figure~\ref{fig:high_dynamic_range}. We use the high-sensitivity recording for our dataset whenever possible. In the few (5\% of the dataset) cases, like yelling, where the high-sensitivity microphone clips, we replace it with the lower-sensitivity microphone. To maintain the same audio characteristics between both microphones, we measure the transfer function between them using a sine-sweep and deconvolution and equalize the low-sensitivity microphone accordingly. See the project page for examples\footnotemark[1].

\textbf{High Speaker Diversity.}
We recorded 107 speakers from diverse demographic backgrounds, each for close to one hour, resulting in a dataset with 100$\,$h of clean speech. Our speakers range from age 18 to 75 and span various ethnicities, including African American, Caucasian, Hispanic, and Asian. Participants are 44\% male, 53\% female, and 3\% non-binary.

\textbf{High Content Diversity.}
Each speaker follows a script that covers a wide variety of speech styles. The script contains a large portion of phonetically balanced sentence reading in seven different styles (regular, loud, whisper, fast, slow, low pitch, and high pitch). Additionally, it contains 18 minutes of conversational freeform speech, where participants freely reply to open-ended questions asked by an operator or talk about vacations, hobbies, or professions. To cover the wide range of emotional speech, we ask participants to read three sentences and describe an image with a specific emotional tone for each of 22 different emotions, including base emotions like ecstasy, fear, anger, or sadness, and nuanced emotions like serenity or adoration. To cover the full variety of human sounds, we additionally include short sections with non-speech sounds like laughter, yelling, or crying, vegetative sounds like coughing or yawning, interjection words, and melodic sounds. A trained operator monitors the participant during the recordings and ensures that speaking styles and prompts are followed as requested and re-record faulty segments.

\section{Benchmarks}

The EARS dataset enables various speech processing tasks to be evaluated in a controlled and comparable way. Here, we present benchmarks for speech enhancement and dereverberation tasks. According to the typical convention, we divide the data into training, validation, and test splits. We select participants \emph{p001} to \emph{p099} for training, \emph{p100} and \emph{p101} as validation speakers, and \emph{p102} to \emph{p107} as test speakers. We use all speech files except utterances containing interjection, melodic, nonverbal, or vegetative sounds. We cut longer files in the validation and training splits every 10$\,$s to be at least 4$\,$s long. For the test set, we provide cutting times and exclude files that are longer than 29$\,$s. This results in 32,485 files (86.8$\,$h) for training, 632 files (1.7$\,$h) for validation, and 886 files (3.7$\,$h) for the test. Data generation scripts can be found online\footnotemark[1]. 

\begin{table}[t]
\footnotesize
\begin{tabularx}{1.0\linewidth}{@{}lcrrr@{}}
\toprule
& pub. date & \#$\,$params & G\acp{MAC} & proc/s [s]\\ \midrule
Conv-TasNet~\cite{luo2019conv} & May 2019 & 8.7$\,$M & 28 & 0.015\\
CDiffuSE~\cite{lu2022conditional} & May 2022 & 18.1$\,$M & 18,382 & 42.268\\
Demucs~\cite{rouard2022hybrid} & June 2023 & 83.6$\,$M & 60 & 0.027\\
SGMSE+ \cite{richter2023speech} & June 2023 & 64.8$\,$M & 47,984 & 2.575 \\ \bottomrule
\end{tabularx}
\caption{\textbf{Baseline methods}. Date of publication, number of parameters, \acp{MAC} for an input of four seconds, and average processing time per one-second input length.}
\label{tab:num_params_macs}
\vspace{-1em}
\end{table}

\begin{table*}[t]
\footnotesize
\centering
\begin{tabularx}{1.0\linewidth}{@{}X|cccc|cc|c@{}}
\toprule
 & \textbf{POLQA} & \textbf{SI-SDR} [dB] & \textbf{PESQ} & \textbf{ESTOI} & \textbf{SIGMOS} & \textbf{DNSMOS} & \textbf{WER} [\%] \\
&\scriptsize(14$\,$kHz)&\scriptsize(24$\,$kHz)&\scriptsize(7$\,$kHz)&\scriptsize(5$\,$kHz)&\scriptsize(24$\,$kHz)&\scriptsize(8$\,$kHz)&\scriptsize(8$\,$kHz)\\
\midrule
Noisy & $1.71 \pm 0.56$ & $5.98 \pm 6.10$ & $1.24 \pm 0.22$ & $0.49 \pm 0.15$ & $1.95 \pm 0.39$ & $2.74 \pm 0.29$ & $33 \pm 29$ \\
\midrule
Conv-TasNet \cite{luo2019conv} & $2.73 \pm 0.78$ & $\mathbf{16.93 \pm 4.36}$ & $2.31 \pm 0.59$ & $0.70 \pm 0.14$ & $2.69 \pm 0.42$ & $3.47 \pm 0.31$ & $20 \pm 20$ \\
CDiffuSE \cite{lu2022conditional} & $1.81 \pm 0.50$ & $8.35 \pm 3.13$ & $1.60 \pm 0.40$ & $0.53 \pm 0.15$ & $2.08 \pm 0.31$ & $2.87 \pm 0.26$ & $32 \pm 27$ \\
Demucs \cite{rouard2022hybrid} & $2.97 \pm 0.75$ & $16.92 \pm 4.35$ & $2.37 \pm 0.58$ & $0.71 \pm 0.14$ & $2.87 \pm 0.43$ & $3.66 \pm 0.30$ & $17 \pm 18$ \\
SGMSE+ \cite{richter2023speech} & $\mathbf{3.40 \pm 0.73}$ & $16.78 \pm 4.47$ & $\mathbf{2.50 \pm 0.62}$ & $\mathbf{0.73 \pm 0.13}$ & $\mathbf{3.41 \pm 0.41}$ & $\mathbf{3.88 \pm 0.26}$ & $\mathbf{16 \pm 18}$ \\
\bottomrule
\end{tabularx}
\caption{\textbf{Results on EARS-WHAM}. Column groups are the method name, intrusive metrics, non-intrusive metrics, and WER. Below each metric is the maximum frequency taken into account for the assessment. Values indicate mean and standard deviation.}
\label{tab:results_ears_wham}
\vspace{-1em}
\end{table*}

\begin{table*}[t]
\footnotesize
\centering
\begin{tabularx}{1.0\linewidth}{@{}X|cccc|cc|c@{}}
\toprule
 & \textbf{POLQA} & \textbf{SI-SDR} [dB] & \textbf{PESQ} & \textbf{ESTOI} & \textbf{SIGMOS} & \textbf{DNSMOS} & \textbf{WER} [\%] \\
\midrule
Noisy & $1.81 \pm 0.60$ & $6.48 \pm 6.76$ & $1.28 \pm 0.32$ & $0.57 \pm 0.18$ & $1.97 \pm 0.44$ & $2.79 \pm 0.37$ & $28 \pm 25$ \\
\midrule
Conv-TasNet \cite{luo2019conv} & $2.68 \pm 0.75$ & $16.56 \pm 5.80$ & $2.41 \pm 0.63$ & $0.75 \pm 0.14$ & $2.70 \pm 0.38$ & $3.43 \pm 0.35$ & $23 \pm 22$ \\
CDiffuSE \cite{lu2022conditional} & $1.93 \pm 0.61$ & $8.22 \pm 3.97$ & $1.64 \pm 0.46$ & $0.59 \pm 0.17$ & $2.09 \pm 0.34$ & $2.92 \pm 0.29$ & $31 \pm 25$ \\
Demucs \cite{rouard2022hybrid} & $3.03 \pm 0.79$ & $\mathbf{16.81 \pm 5.94}$ & $2.50 \pm 0.63$ & $0.76 \pm 0.14$ & $2.82 \pm 0.43$ & $3.62 \pm 0.34$ & $19 \pm 20$ \\
SGMSE+ \cite{richter2023speech} & $\mathbf{3.35 \pm 0.82}$ & $16.43 \pm 6.12$ & $\mathbf{2.59 \pm 0.70}$ & $\mathbf{0.78 \pm 0.13}$ & $\mathbf{3.30 \pm 0.40}$ & $\mathbf{3.79 \pm 0.30}$ & $\mathbf{19 \pm 19}$ \\
\bottomrule
\end{tabularx}
\caption{\textbf{Results for the blind test}. Column groups are the method name, intrusive metrics, non-intrusive metrics, and WER. Values indicate mean and standard deviation.}
\label{tab:results_ears_blind}
\vspace{-1em}
\end{table*}

\begin{table*}[t]
    \footnotesize
\begin{tabularx}{1.0\linewidth}{@{}lXccccXcccc}
\toprule
& & \multicolumn{4}{c}{\textbf{(a) POLQA}} & & \multicolumn{4}{c}{\textbf{(b) SI-SDR} [dB]} \\
\cmidrule(lr){3-6} \cmidrule(lr){8-11} & & 0$\,$dB & 5$\,$dB & 10$\,$dB & 15$\,$dB & & 0$\,$dB & 5$\,$dB & 10$\,$dB & 15$\,$dB  \\ \midrule
Noisy &  & $1.2 \pm 0.2$ & $1.4 \pm 0.3$ & $1.9 \pm 0.4$ & $2.4 \pm 0.4$ &  & $-1.6 \pm 2.4$ & $3.5 \pm 2.2$ & $8.6 \pm 2.6$ & $13.5 \pm 2.2$ \\
\midrule
Conv-TasNet \cite{luo2019conv} &  & $1.9 \pm 0.5$ & $2.5 \pm 0.5$ & $3.1 \pm 0.5$ & $3.5 \pm 0.5$ &  & $11.7 \pm 2.6$ & $15.2 \pm 2.0$ & $\mathbf{19.0 \pm 2.0}$ & $\mathbf{21.8 \pm 1.7}$ \\
CDiffuSE \cite{lu2022conditional} &  & $1.3 \pm 0.2$ & $1.6 \pm 0.3$ & $2.0 \pm 0.4$ & $2.3 \pm 0.5$ &  & $4.5 \pm 1.9$ & $8.1 \pm 1.9$ & $10.1 \pm 2.0$ & $10.7 \pm 2.1$ \\
Demucs \cite{rouard2022hybrid} &  & $2.1 \pm 0.5$ & $2.7 \pm 0.5$ & $3.4 \pm 0.4$ & $3.7 \pm 0.3$ &  & $\mathbf{11.9 \pm 2.6}$ & $\mathbf{15.3 \pm 2.3}$ & $18.9 \pm 2.3$ & $21.7 \pm 2.1$ \\
SGMSE+ \cite{richter2023speech} &  & $\mathbf{2.6 \pm 0.6}$ & $\mathbf{3.3 \pm 0.6}$ & $\mathbf{3.8 \pm 0.3}$ & $\mathbf{4.0 \pm 0.3}$ &  & $11.6 \pm 2.7$ & $15.1 \pm 2.2$ & $18.8 \pm 2.2$ & $21.8 \pm 2.0$ \\
\bottomrule
\end{tabularx}
\vskip 1em
\begin{tabularx}{1.0\linewidth}{@{}lXccccXcccc}
& & \multicolumn{4}{c}{\textbf{(c) ESTOI}} & & \multicolumn{4}{c}{\textbf{(d) WER} [\%]} \\
\cmidrule(lr){3-6} \cmidrule(lr){8-11} & & 0$\,$dB & 5$\,$dB & 10$\,$dB & 15$\,$dB & & 0$\,$dB & 5$\,$dB & 10$\,$dB & 15$\,$dB  \\ \midrule
Noisy &  & $0.32 \pm 0.08$ & $0.44 \pm 0.09$ & $0.56 \pm 0.10$ & $0.65 \pm 0.11$ &  & $63 \pm 25$ & $39 \pm 24$ & $18 \pm 17$ & $12 \pm 16$ \\
\midrule
Conv-TasNet \cite{luo2019conv} &  & $0.58 \pm 0.14$ & $0.67 \pm 0.12$ & $0.75 \pm 0.11$ & $0.79 \pm 0.10$ &  & $39 \pm 22$ & $21 \pm 18$ & $11 \pm 12$ & $8 \pm 12$ \\
CDiffuSE \cite{lu2022conditional} &  & $0.37 \pm 0.10$ & $0.50 \pm 0.11$ & $0.60 \pm 0.11$ & $0.65 \pm 0.11$ &  & $61 \pm 24$ & $36 \pm 22$ & $18 \pm 16$ & $14 \pm 18$ \\
Demucs \cite{rouard2022hybrid} &  & $0.60 \pm 0.13$ & $0.69 \pm 0.11$ & $0.76 \pm 0.10$ & $0.80 \pm 0.10$ &  & $32 \pm 21$ & $\mathbf{17 \pm 16}$ & $9 \pm 11$ & $7 \pm 11$ \\
SGMSE+ \cite{richter2023speech} &  & $\mathbf{0.63 \pm 0.13}$ & $\mathbf{0.72 \pm 0.11}$ & $\mathbf{0.78 \pm 0.10}$ & $\mathbf{0.81 \pm 0.10}$ &  & $\mathbf{30 \pm 21}$ & $\mathbf{17 \pm 16}$ & $\mathbf{8 \pm 9}$ & $\mathbf{6 \pm 10}$ \\
\bottomrule
\end{tabularx}
    \caption{\textbf{Results per input SNR}. Mean and standard deviation for (a) POLQA, (b) SI-SDR, (c) ESTOI, and (d) WER on EARS-WHAM.}
    \label{tab:results_per_snr}
    \vspace{-1em}
\end{table*}

\begin{table*}[t]
\footnotesize
\centering
\begin{tabularx}{1.0\linewidth}{@{}X|cccc|cc|c@{}}
\toprule
 & \textbf{POLQA} & \textbf{SI-SDR} [dB] & \textbf{PESQ} & \textbf{ESTOI} & \textbf{SIGMOS} & \textbf{MOS Reverb} & \textbf{WER} [\%] \\
\midrule
Reverberant & $1.75 \pm 0.48$ & $-16.17 \pm 9.77$ & $1.48 \pm 0.37$ & $0.52 \pm 0.17$ & $2.77 \pm 0.43$ & $2.99 \pm 0.74$ & $25 \pm 25$ \\
\midrule
SGMSE+ \cite{richter2023speech} & $\mathbf{3.61 \pm 0.63}$ & $\mathbf{5.79 \pm 7.97}$ & $\mathbf{3.03 \pm 0.67}$ & $\mathbf{0.85 \pm 0.09}$ & $\mathbf{3.49 \pm 0.43}$ & $\mathbf{4.73 \pm 0.21}$ & $\mathbf{9 \pm 12}$ \\
\bottomrule
\end{tabularx}
\caption{\textbf{Results on EARS-Reverb}. Column groups are the method name, intrusive metrics, non-intrusive metrics, and WER in percent. Values indicate mean and standard deviation.}
\label{tab:results_ears_reverb}
\vspace{-1em}
\end{table*}

\subsection{EARS-WHAM}

For the speech enhancement task, we construct the EARS-WHAM dataset, which mixes speech from the EARS~dataset with real noise recordings from the WHAM! dataset~\cite{wichern19_interspeech} (CC BY-NC 4.0 license). We mix speech and noise files at \acp{SNR} randomly sampled in a range of $[-2.5, 17.5]\,$dB, where we compute the \ac{SNR} using \ac{LKFS} standardized in ITU-R BS.1770~\cite{itu-r-bs.1770-5} to obtain a more perceptually meaningful scaling and also to remove silent regions from the \ac{SNR} computation~\cite{steinmetz2021pyloudnorm}. We additionally create a \textbf{blind test set} for which we only publish the noisy audio files but not the clean ground truth. It contains 743 files (2$\,$h) from six speakers (3 male, 3 female) that are not part of the EARS dataset and noise especially recorded for this test set. We set up an evaluation server for blind evaluation on this test set, which can be found online\footnotemark[1].

\subsection{EARS-Reverb}

For the task of dereverberation, we use real recorded \acp{RIR} from multiple public datasets \cite{eaton2016estimation, jeub2009binaural, prawda2022robust, fejgin2023brudex, carlo2021dechorate, amengual2020open, Palimpsest} (CC BY 4.0, MIT license). Table \ref{tab:rir_datasets} shows statistics on the \ac{RIR} datasets used. All \acp{RIR} are fullband, and we use a randomly selected channel for multi-channel recordings. We generate reverberant speech by convolving the clean speech with the \ac{RIR}. To avoid a time delay between the reverberant and clean speech signal caused by the direct path of the \ac{RIR}, we cut off the beginning of the \ac{RIR} up to the index with the highest amplitude. We only use \acp{RIR} with an $\text{RT}_{60}$ reverberation time that does not exceed 2$\,$s. Finally, we normalize the loudness of the reverberant speech to the loudness of the clean speech using \ac{LKFS}.

\section{Baselines and Evaluation}

\subsection{Baselines}

Table~\ref{tab:num_params_macs} shows all baseline methods with the date of publication, number of parameters, \acfp{MAC} for an input of 4$\,$s using the ptflops package\footnote{\texttt{https://pypi.org/project/ptflops/}}, and the processing time per input second. We calculate the processing time per second averaged over 20 utterances from the test set using an NVIDIA RTX A6000 \ac{GPU}. 

\textbf{Conv-TasNet}~\cite{luo2019conv} is a predictive method initially proposed for speech separation that operates in the time domain. Identical to the original approach, we learn 2$\,$ms filters, which correspond to kernels of size 120 and stride of 60 at a sampling rate of 48$\,$kHz. We train with a batch size of 4 using one \ac{GPU}.

\textbf{CDiffuSE}~\cite{lu2022conditional} is a generative speech enhancement method based on a conditional diffusion process defined in the time domain. We adapt the method for 48$\,$kHz by using a 3072-point \ac{STFT}, resulting in 1537 frequency bins for the conditioner. We train the large model with a batch size of 16 using two \acp{GPU}.

\textbf{Demucs v4}~\cite{rouard2022hybrid} is a predictive model originally proposed for music separation. We train with batch size 8 using one \ac{GPU}.

\textbf{SGMSE+}~\cite{richter2023speech} is a generative speech enhancement method based on a conditional diffusion process defined in the complex \ac{STFT} domain. We adapt the method for 48$\,$kHz by using 1534-point \ac{STFT} with hop size 384. We use $\alpha=0.667$ and $\beta=0.065$ for the \ac{STFT} amplitude compression and $\sigma_\text{min}=0.1$, $\sigma_\text{max}=1$, and $\gamma=2$ for the stochastic differential equation. We train with a batch size of 4 using four \acp{GPU}.

\subsection{Metrics}

We employ \emph{intrusive} metrics that rate the processed signal in relation to the clean reference signal and \emph{non-intrusive} metrics, which assess the performance only using the processed signal.

Intrusive metrics include the \ac{POLQA} \cite{polqa2018} for predicting speech quality, which takes values from 1 (poor) to 5 (excellent) as usual for \ac{MOS}. We report the \ac{PESQ}~\cite{rixPerceptualEvaluationSpeech2001}, which is the predecessor of POLQA and is still widely used in the research community. The PESQ score lies between 1 (poor) and 4.5 (excellent). We further use \ac{ESTOI}~\cite{jensen2016algorithm} as an intrusive measure of speech intelligibility. This metric yields values between 0 and 1, with higher values indicating better intelligibility. Moreover, we calculate the \ac{SI-SDR}~\cite{leroux2018sdr} measured in dB, with higher values indicating better performance.

Non-intrusive metrics include the SIGMOS estimator~\cite{ristea2024icassp}, which is a speech quality assessment model based on a multi-dimensional listening test~\cite{naderi2024multi}. We report the overall quality (SIGMOS) and the reverberation assessment (MOS Reverb, only in Table \ref{tab:results_ears_reverb}). In addition, we use the speech quality assessment model DNSMOS \cite{reddy2021dnsmos} that is trained on human ratings obtained from listening experiments based on ITU-T P.808 \cite{itu-t-rec808}.

To evaluate the effect of speech enhancement on \ac{ASR}, we use Quartz\-Net\-15\-x5Base-En from the NeMo toolkit \cite{kuchaiev2019nemo} as a downstream \ac{ASR} system and report the \acf{WER}. We obtain the reference transcriptions by performing \ac{ASR} on the clean speech utterances. 

\subsection{Evaluation}

We provide an empirical evaluation of the speech enhancement and dereverberation benchmarks.
Listening examples for both tasks can be found online\footnotemark[1].

\begin{figure}[t]
    \centering
    \includegraphics[width=0.8\linewidth]{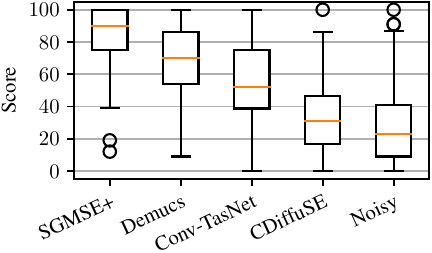}
     \caption{\textbf{Results of the listening test}. Subjective scores based on 20 participants visualized in a standard box plot.}
    \label{fig:listening_test}
    \vspace{-1em}
\end{figure}

\textbf{Speech Enhancement.}
In Table~\ref{tab:results_ears_wham} and Table~\ref{tab:results_ears_blind}, we show speech enhancement results on the EARS-WHAM test set and the blind test set, respectively. Among the methods, the generative SGMSE+~\cite{richter2023speech} performs the best across most metrics, with particularly high scores in POLQA and SIGMOS.
Demucs~\cite{rouard2022hybrid}, as a representative of predictive methods, convinces with strong results, too, although falling slightly behind SGMSE+.

\textbf{Listening Test.}
We conduct a MUSHRA-like (\emph{Multiple Stimuli with Hidden Reference and Anchor}) listening test on EARS-WHAM with 20 participants. We randomly sample 10 distinct utterances from the test set in a gender-balanced way (5 male, 5 female). We use the clean audio as the hidden reference and the noisy audio as the hidden anchor. As stimuli, we use enhanced files of each noisy utterance from the four methods. We present participants with six audio files (four stimuli, hidden reference, hidden anchor) per utterance and ask them to \enquote{rate the overall quality considering artifacts and residual noise} of each on a scale of 0--100. The trends support the quantitative evaluation, demonstrating that SGMSE+~\cite{richter2023speech} is the preferred approach, closely followed by Demucs~\cite{rouard2022hybrid}, see Figure~\ref{fig:listening_test}.

\textbf{Effect of Input SNR.}
Table~\ref{tab:results_per_snr} shows POLQA, SI-SDR, ESTOI, and WER scores segmented by input SNR, where 0$\,$dB denotes the range $[-2.5, 2.5]\,$dB, and each subsequent 5$\,$dB increment representing the next range. As expected, there is a trend for better performance at higher input SNR, as well as smaller standard deviations than on the full test set.

\textbf{Effect of Speaking Style and Emotion.}
We compare the performance of all baseline methods with respect to speaking style and selected core emotions in Table~\ref{tab:speaking_style} and~\ref{tab:emotions}. We observe worse performance for whispered speech, which is expected since such voiceless speech is particularly difficult to recover after contamination with noise. Furthermore, it can be seen that all considered approaches trained on EARS-WHAM generalize well to emotional speech.

\textbf{Dereverberation.}
Blind dereverberation with only a single microphone is known to be challenging, and recent results suggest that generative approaches are particularly well suited for this task \cite{lemercier2023icassp}. In Table~\ref{tab:results_ears_reverb}, we show dereverberation results on the EARS-Reverb test set, using the diffusion-based generative model SGMSE+ \cite{richter2023speech}.

\begin{table}[t]
\footnotesize
\centering
\begin{tabularx}{1.0\linewidth}{@{}Xccccc@{}}
\toprule
 & regular & whisper & loud & slow & fast \\
\midrule
Noisy & $1.74$ & $1.85$ & $1.75$ & $1.68$ & $1.72$ \\
\midrule
Conv-TasNet \cite{luo2019conv} & $2.82$ & $2.49$ & $2.79$ & $2.75$ & $3.17$ \\
CDiffuSE \cite{lu2022conditional} & $1.78$ & $1.68$ & $1.93$ & $1.72$ & $2.02$ \\
Demucs \cite{rouard2022hybrid} & $2.95$ & $2.82$ & $3.10$ & $2.86$ & $3.27$ \\
SGMSE+ \cite{richter2023speech} & $3.39$ & $2.89$ & $3.70$ & $3.33$ & $3.64$ \\
\bottomrule
\end{tabularx}
\caption{\textbf{POLQA for different speaking styles}. Mean values.}
\vspace{-1em}
\label{tab:speaking_style}
\end{table}

\begin{table}[t]
\footnotesize
\centering
\begin{tabularx}{1.0\linewidth}{@{}Xcccccc@{}}
\toprule
 & neutral & anger & desire & pain & relief \\
\midrule
Noisy & $1.77$ & $1.61$ & $1.65$ & $1.86$ & $1.74$ \\
\midrule
Conv-TasNet \cite{luo2019conv} & $2.78$ & $2.53$ & $2.71$ & $2.72$ & $2.76$ \\
CDiffuSE \cite{lu2022conditional} & $1.65$ & $1.94$ & $1.76$ & $1.94$ & $1.78$ \\
Demucs \cite{rouard2022hybrid} & $2.93$ & $2.91$ & $2.82$ & $3.02$ & $2.91$ \\
SGMSE+ \cite{richter2023speech} & $3.18$ & $3.45$ & $3.15$ & $3.40$ & $3.28$ \\
\bottomrule
\end{tabularx}
\caption{\textbf{POLQA for different emotions}. Mean values.}
\label{tab:emotions}
\vspace{-1em}
\end{table}

\section{Conclusion}

We released EARS, a dataset with high speaker and speaking style diversity spanning the full range of human speech. We hope this dataset will serve the community as a useful source to tackle new frontiers in speech processing. We additionally provided a speech enhancement and dereverberation benchmark on this new large-scale dataset and compared predictive and generative baselines to set a standard for future speech enhancement work on EARS.

\section{Acknowledgements}

This work has been funded by the German Research Foundation (DFG) in the transregio project Crossmodal Learning (TRR 169) and DASHH (Data Science in Hamburg -- Helmholtz Graduate School for the Structure of Matter) with Grant-No. HIDSS-0002. We would like to thank J. Berger and Rohde \& Schwarz SwissQual AG for their support with POLQA.

\bibliographystyle{IEEEtran}
\bibliography{refs}

\begin{thebibliography}{10}
\providecommand{\url}[1]{#1}
\csname url@samestyle\endcsname
\providecommand{\newblock}{\relax}
\providecommand{\bibinfo}[2]{#2}
\providecommand{\BIBentrySTDinterwordspacing}{\spaceskip=0pt\relax}
\providecommand{\BIBentryALTinterwordstretchfactor}{4}
\providecommand{\BIBentryALTinterwordspacing}{\spaceskip=\fontdimen2\font plus
\BIBentryALTinterwordstretchfactor\fontdimen3\font minus
  \fontdimen4\font\relax}
\providecommand{\BIBforeignlanguage}[2]{{%
\expandafter\ifx\csname l@#1\endcsname\relax
\typeout{** WARNING: IEEEtran.bst: No hyphenation pattern has been}%
\typeout{** loaded for the language `#1'. Using the pattern for}%
\typeout{** the default language instead.}%
\else
\language=\csname l@#1\endcsname
\fi
#2}}
\providecommand{\BIBdecl}{\relax}
\BIBdecl

\bibitem{mohamed2022self}
A.~Mohamed, H.-y. Lee, L.~Borgholt, J.~D. Havtorn, J.~Edin, C.~Igel,
  K.~Kirchhoff, S.-W. Li, K.~Livescu, L.~Maal{\o}e \emph{et~al.},
  ``Self-supervised speech representation learning: A review,'' \emph{IEEE
  Journal of Selected Topics in Signal Processing}, vol.~16, no.~6, pp.
  1179--1210, 2022.

\bibitem{tan2021survey}
X.~Tan, T.~Qin, F.~Soong, and T.-Y. Liu, ``A survey on neural speech
  synthesis,'' \emph{arXiv preprint arXiv:2106.15561}, 2021.

\bibitem{wang2018supervised}
D.~Wang and J.~Chen, ``Supervised speech separation based on deep learning: An
  overview,'' \emph{IEEE/ACM Transactions on Audio, Speech, and Language
  Processing}, vol.~26, no.~10, pp. 1702--1726, 2018.

\bibitem{panayotov2015librispeech}
V.~Panayotov, G.~Chen, D.~Povey, and S.~Khudanpur, ``Librispeech: An {ASR}
  corpus based on public domain audio books,'' in \emph{IEEE International
  Conference on Acoustics, Speech and Signal Processing}, 2015, pp. 5206--5210.

\bibitem{yamagishi2019vctk}
\BIBentryALTinterwordspacing
J.~Yamagishi, C.~Veaux, and K.~MacDonald, ``{CSTR VCTK} corpus: English
  multi-speaker corpus for {CSTR} voice cloning toolkit (version 0.92),'' 2019.
  [Online]. Available: \url{https://datashare.ed.ac.uk/handle/10283/3443}
\BIBentrySTDinterwordspacing

\bibitem{luo2019conv}
Y.~Luo and N.~Mesgarani, ``{Conv-TasNet}: Surpassing ideal time--frequency
  magnitude masking for speech separation,'' \emph{IEEE/ACM Transactions on
  Audio, Speech, and Language Processing}, vol.~27, no.~8, pp. 1256--1266,
  2019.

\bibitem{rouard2022hybrid}
S.~Rouard, F.~Massa, and A.~D{\'e}fossez, ``Hybrid transformers for music
  source separation,'' in \emph{IEEE International Conference on Acoustics,
  Speech and Signal Processing}, 2023.

\bibitem{lu2022conditional}
Y.-J. Lu, Z.-Q. Wang, S.~Watanabe, A.~Richard, C.~Yu, and Y.~Tsao,
  ``Conditional diffusion probabilistic model for speech enhancement,'' in
  \emph{IEEE International Conference on Acoustics, Speech and Signal
  Processing}, 2022, pp. 7402--7406.

\bibitem{richter2023speech}
J.~Richter, S.~Welker, J.-M. Lemercier, B.~Lay, and T.~Gerkmann, ``Speech
  enhancement and dereverberation with diffusion-based generative models,''
  \emph{IEEE/ACM Transactions on Audio, Speech, and Language Processing},
  vol.~31, pp. 2351--2364, 2023.

\bibitem{dubey2023icassp}
H.~Dubey, A.~Aazami, V.~Gopal, B.~Naderi, S.~Braun, R.~Cutler, H.~Gamper,
  M.~Golestaneh, and R.~Aichner, ``{ICASSP} 2023 deep noise suppression
  challenge,'' in \emph{IEEE International Conference on Acoustics, Speech and
  Signal Processing}, 2023.

\bibitem{Martinez-Lucas_2020}
L.~Martinez-Lucas, M.~Abdelwahab, and C.~Busso, ``The {MSP}-conversation
  corpus,'' in \emph{ISCA Interspeech}, 2020, pp. 1823--1827.

\bibitem{Lotfian_2019_3}
R.~Lotfian and C.~Busso, ``Building naturalistic emotionally balanced speech
  corpus by retrieving emotional speech from existing podcast recordings,''
  \emph{IEEE Transactions on Affective Computing}, vol.~10, no.~4, pp.
  471--483, 2019.

\bibitem{ljspeech17}
\BIBentryALTinterwordspacing
K.~Ito and L.~Johnson, ``The {LJ Speech Dataset},'' 2017. [Online]. Available:
  \url{https://keithito.com/LJ-Speech-Dataset/}
\BIBentrySTDinterwordspacing

\bibitem{garofolo1993timit}
J.~S. Garofolo, ``{TIMIT} acoustic phonetic continuous speech corpus,''
  \emph{Linguistic Data Consortium}, 1993.

\bibitem{garofolo1996wsj0}
\BIBentryALTinterwordspacing
J.~S. Garofolo, D.~Graff, D.~Paul, and D.~Pallett, ``{CSR-I (WSJ0) Complete -
  Linguistic Data Consortium},'' 1993. [Online]. Available:
  \url{https://catalog.ldc.upenn.edu/LDC93s6a}
\BIBentrySTDinterwordspacing

\bibitem{eaton2016estimation}
J.~Eaton, N.~D. Gaubitch, A.~H. Moore, and P.~A. Naylor, ``Estimation of room
  acoustic parameters: The {ACE} challenge,'' \emph{IEEE/ACM Transactions on
  Audio, Speech, and Language Processing}, vol.~24, no.~10, pp. 1681--1693,
  2016.

\bibitem{jeub2009binaural}
M.~Jeub, M.~Schafer, and P.~Vary, ``A binaural room impulse response database
  for the evaluation of dereverberation algorithms,'' in \emph{IEEE Int.
  Conference on Digital Signal Processing}, 2009.

\bibitem{prawda2022robust}
K.~Prawda, S.~J. Schlecht, and V.~V{\"a}lim{\"a}ki, ``Robust selection of clean
  swept-sine measurements in non-stationary noise,'' \emph{The Journal of the
  Acoustical Society of America}, vol. 151, no.~3, pp. 2117--2126, 2022.

\bibitem{fejgin2023brudex}
D.~Fejgin, W.~Middelberg, and S.~Doclo, ``{BRUDEX} database: Binaural room
  impulse responses with uniformly distributed external microphones,'' in
  \emph{Proc. ITG Conference on Speech Communication}, 2023, pp. 126--130.

\bibitem{carlo2021dechorate}
D.~D. Carlo, P.~Tandeitnik, C.~Foy, N.~Bertin, A.~Deleforge, and S.~Gannot,
  ``{dEchorate}: a calibrated room impulse response dataset for echo-aware
  signal processing,'' \emph{EURASIP Journal on Audio, Speech, and Music
  Processing}, 2021.

\bibitem{amengual2020open}
S.~V. Amengual~Gari, B.~Sahin, D.~Eddy, and M.~Kob, ``Open database of spatial
  room impulse responses at {Detmold} university of music,'' in \emph{Audio
  Engineering Society Convention 149}, 2020.

\bibitem{Palimpsest}
\BIBentryALTinterwordspacing
``A sonic {Palimpsest}: Revisiting {Chatham} historic dockyards.'' [Online].
  Available:
  \url{https://research.kent.ac.uk/sonic-palimpsest/impulse-responses/}
\BIBentrySTDinterwordspacing

\bibitem{wichern19_interspeech}
G.~Wichern, J.~Antognini, M.~Flynn, L.~R. Zhu, E.~McQuinn, D.~Crow, E.~Manilow,
  and J.~L. Roux, ``{WHAM!}: Extending speech separation to noisy
  environments,'' in \emph{ISCA Interspeech}, 2019, pp. 1368--1372.

\bibitem{itu-r-bs.1770-5}
\BIBentryALTinterwordspacing
{Recommendation ITU-R BS.1770-5}, ``Algorithms to measure audio programme
  loudness and true-peak audio level,'' \emph{International Telecommunication
  Union ({ITU})}, 2023. [Online]. Available:
  \url{https://www.itu.int/rec/R-REC-BS.1770-5-202311-I/en}
\BIBentrySTDinterwordspacing

\bibitem{steinmetz2021pyloudnorm}
C.~J. Steinmetz and J.~D. Reiss, ``pyloudnorm: {A} simple yet flexible loudness
  meter in python,'' in \emph{150th AES Convention}, 2021.

\bibitem{polqa2018}
\BIBentryALTinterwordspacing
{ITU-T Rec. P.863}, ``Perceptual objective listening quality prediction,''
  \emph{Int. Telecom. Union ({ITU})}, 2018. [Online]. Available:
  \url{https://www.itu.int/rec/T-REC-P.863-201803-I/en}
\BIBentrySTDinterwordspacing

\bibitem{rixPerceptualEvaluationSpeech2001}
A.~Rix, J.~Beerends, M.~Hollier, and A.~Hekstra, ``Perceptual evaluation of
  speech quality ({{PESQ}}) - a new method for speech quality assessment of
  telephone networks and codecs,'' in \emph{IEEE International Conference on
  Acoustics, Speech and Signal Processing}, 2001, pp. 749--752.

\bibitem{jensen2016algorithm}
J.~Jensen and C.~H. Taal, ``An algorithm for predicting the intelligibility of
  speech masked by modulated noise maskers,'' \emph{IEEE/ACM Transactions on
  Audio, Speech, and Language Processing}, vol.~24, no.~11, pp. 2009--2022,
  2016.

\bibitem{leroux2018sdr}
J.~Le~Roux, S.~Wisdom, H.~Erdogan, and J.~R. Hershey, ``{SDR}--half-baked or
  well done?'' in \emph{IEEE International Conference on Acoustics, Speech and
  Signal Processing}, 2019, pp. 626--630.

\bibitem{ristea2024icassp}
N.~C. Ristea, A.~Saabas, R.~Cutler, B.~Naderi, S.~Braun, and S.~Branets,
  ``{ICASSP} 2024 speech signal improvement challenge,'' in \emph{IEEE
  International Conference on Acoustics, Speech and Signal Processing}, 2024.

\bibitem{naderi2024multi}
B.~Naderi, R.~Cutler, and N.-C. Ristea, ``Multi-dimensional speech quality
  assessment in crowdsourcing,'' in \emph{IEEE International Conference on
  Acoustics, Speech and Signal Processing}, 2024.

\bibitem{reddy2021dnsmos}
C.~K. Reddy, V.~Gopal, and R.~Cutler, ``{DNSMOS}: A non-intrusive perceptual
  objective speech quality metric to evaluate noise suppressors,'' in
  \emph{IEEE International Conference on Acoustics, Speech and Signal
  Processing}, 2021, pp. 6493--6497.

\bibitem{itu-t-rec808}
\BIBentryALTinterwordspacing
{ITU-T Rec. P.808}, ``Subjective evaluation of speech quality with a
  crowdsourcing approach,'' \emph{International Telecommunication Union}, 2021.
  [Online]. Available: \url{https://www.itu.int/rec/T-REC-P.808-202106-I/en}
\BIBentrySTDinterwordspacing

\bibitem{kuchaiev2019nemo}
O.~Kuchaiev, J.~Li, H.~Nguyen, O.~Hrinchuk, R.~Leary, B.~Ginsburg, S.~Kriman,
  S.~Beliaev, V.~Lavrukhin, J.~Cook \emph{et~al.}, ``{NeMo}: a toolkit for
  building {AI} applications using neural modules,'' \emph{arXiv preprint
  arXiv:1909.09577}, 2019.

\bibitem{lemercier2023icassp}
J.-M. Lemercier, J.~Richter, S.~Welker, and T.~Gerkmann, ``Analysing
  diffusion-based generative approaches versus discriminative approaches for
  speech restoration,'' in \emph{IEEE International Conference on Acoustics,
  Speech and Signal Processing}, 2023.

\end{thebibliography}

\end{document}